# Strongly Magnetically Tuned Coupling Strength and Nonlinearity in CrSBr Exciton-Polaritons

Chun Li,[1,†] Chao Shen,[2,3,†] Xuekai Ma,[4] Kwok Kwan Tang,[1] Nai Jiang,[2] Xinyi Deng,[1] Qing Wan,[1,6] Jiepeng Song,[1] Jiaqi Guo,[2] Tian Lan,[1] Feng Li,[7] Yilin Wang,[6] Xinfeng Liu,[5] and Qing Zhang[1,*]

[1]School of Materials Science and Engineering, Peking University, Beijing 100871, China

[2]State Key Laboratory for Superlattices and Microstructures, Institute of Semiconductors, Chinese Academy of Sciences, Beijing 100083, China

[3]Center of Materials Science and Optoelectronics Engineering, University of Chinese Academy of Sciences, Beijing 100049, China

[4]Department of Physics and Center for Optoelectronics and Photonics Paderborn (CeOPP), Paderborn University, Warburger Strasse 100, 33098 Paderborn, Germany

[5]CAS Key Laboratory of Standardization and Measurement for Nanotechnology, CAS Center of Excellence for Nanoscience, National Center for Nanoscience and Technology, Beijing 100190, China

[6]School of Integrated Circuits, Shandong Technology Center of Nanodevices and Integration, State Key Laboratory of Crystal Materials, Shandong University, Jinan 250100, China

[7]Key Laboratory for Physical Electronics and Devices of the Ministry of Education & Shaanxi Key Laboratory of Information Photonic Technique, School of Electronic Science and Engineering, Faculty of Electronic and Information Engineering, Xi'an Jiaotong University, Xi'an 710049, China

[*]Q_zhang@pku.edu.cn

[†]These authors contributed equally to this work.

**Abstract**. Two-dimensional van der Waals (vdW) magnetic semiconductors CrSBr offer an ideal platform to achieve exciton-polaritons correlated with magnetic orders for developing solid-state quantum, spintronic, and photonic devices. However, for the exciton-polaritons formed by lower-energy excitons ($X_L \approx 1.37$ eV), the coupling strength and nonlinear optical response are almost inert to the external magnetic field. Here, we demonstrate robust strong coupling between higher-energy excitons ($X_H \approx 1.8$ eV) and photons that persists up to room temperature, along with giant magnetic-field tunability. The Rabi splitting energy is tuned up to 100 meV within a moderate 0.45 T in-plane magnetic field due to changes in excitonic states during the spin transitions. Besides, we observe significantly enhanced polariton nonlinearity in the intermediate magnetic phase, which exhibits a distinct mode-number dependence and originates from magnon-assisted long-range attractive interactions and coupling strength reduction. These results advance the development of on-demand polariton platforms for spin-correlated quantum optoelectronics.

*Introduction*—In semiconductor excitonic systems, strong exciton-photon coupling gives rise to hybrid exciton-polaritons when the coupling strengths exceed the decay rates of both excitons and photons [1-3], holding significant promise for numerous technological applications such as low-threshold polariton lasers, all-optical switches, and quantum simulators [4-20]. Magnetic fields enable real-time, non-invasive tuning of exciton-polariton properties, including energy dispersion, light-matter coupling strength, and nonlinear response characteristics [21-23]. Moreover, the interaction between exciton-polaritons and magnons facilitates the transfer of magnonic and spintronic information through polariton states, offering distinct advantages for optical detection and information transmission applications [24-28]. To meet this demand, magnetic vdW semiconductors have attracted widespread interest as a bridge connecting strong light-matter interaction and magnetic excitations [29-33]. The vdW antiferromagnet $NiPS_3$ exhibits strong coupling between magnetic excitons and photons, yet the Rabi splitting is surprisingly small, measuring merely a few meV [34]. While pioneering works achieved strong coupling of photons and lower-energy excitons ($X_L$, ~1.37 eV, the first-band transition at Γ point, [Fig.1(a)]) in CrSBr, the magnetic field tuning capability remained restricted to ~15 meV-scale exciton-polariton energy shifts [35-37]. The higher-energy exciton ($X_H$, ~1.8 eV) in CrSBr, arising from second-band transitions at Γ point [38], shows stronger magneto-exciton coupling than the $X_L$ due to enhanced delocalization [39-41]. Its strong exciton-photon coupling behavior under magnetic fields represents a critical experimental platform for polariton physics, though this phenomenon remains unexplored to date.

Here, we report the observation of strong magnetic field engineering strong coupling between $X_H$ excitons and photons in CrSBr crystals. We find remarkably large tuning of the Rabi splitting energy (up to 100 meV) during the antiferromagnetic (AFM) to ferromagnetic (FM) order change. Temperature can mediate the coupled exciton-polariton behaviors by manipulating the interactions between excitons and magnons/phonons. Further, the polariton nonlinearity is stronger in the intermediate magnetic (IM) phase than in the AFM and FM phases, indicating that spin order disruption enhances the attractive exciton-exciton interaction and coupling strength reduction assisted by magnons. These results pave the way for achieving controllable strong coupling in polaritonic optoelectronics and quantum applications.

*Results*—Mesoscopic CrSBr crystals, with thicknesses ranging from tens to hundreds of nanometers, were prepared using mechanical exfoliation and transferred onto the $SiO_2$/Si substrate. The elongated morphology of the crystals reflects the crystallographic anisotropy [Fig. 1(b)]. To probe the excitonic resonance, two types of optical contrasts from differential reflectance spectroscopy and magnetic circular dichroism (MCD) spectroscopy were investigated. Polarization-resolved differential reflectance spectra at normal incidence of the 354 nm-thick CrSBr crystal indicate a reflectance dip at 1.764 eV as the polarization of the reflected light is along the *a*-axis, assigned to an uncoupled cavity mode [Fig. 1(c)]. Due to the anisotropic refractive index, the corresponding uncoupled cavity mode along the *b*-axis is located at ~2.11 eV with a mode number $m = 4$. In contrast, when the polarization is along the *b*-axis, a series of reflectance dips below 1.8 eV varying with the crystal thickness exist (see Fig. S1 [42]), which may be due to exciton-polaritons with different *m*. The calculated refractive index along the *b*-axis according to the reflectance dip energy sharply increases as energy approaches 1.8 eV (see Fig. S2 [42]). This obvious increase exceeds the refractive index variation described by the Sellmeier equation, extracted from the data polarized along the *a*-axis (without exciton effect), indicating the occurrence of strong exciton-photon coupling. The fitted $X_H$ exciton oscillator strength using the Lorentz model for the dielectric function considering magnon effect (see Note S1 [42]) reaches 7.28 $(eV)^2$, which is around four times greater than that of the $X_L$ exciton [35-37]. The larger oscillator strength of $X_H$ excitons can be also identified by fitting differential reflectance spectra (see Fig. S3 [42]), which is due to the larger exciton energy [2]. Additionally, we used a coupled oscillator model to quantitatively fit the energy-detuning [Fig. 1(d)] and energy-wavevector dispersions (see Fig. S4 [42]) extracted from the thickness dependence of differential reflectance spectra, which show anti-crossing features. The fitted average Rabi splitting energy reaches 695 meV, much larger than the average dissipation energy of excitons (~50 meV) and photons (~50 meV), further confirming the formation of exciton-polaritons. Two representative exciton-polariton signals at 1.654 eV and 1.593 eV can be identified in both the differential reflectance and MCD spectra at 80 K, providing a consistency check (see Fig. S5 [42]). The polarization nature of the uncoupled cavity mode and coupled exciton-polaritons is independent of temperature and the magnetic field (see Figs. S5 and S6 [42]), consistent with the reported one-dimensional $X_H$ exciton behavior [39,43].

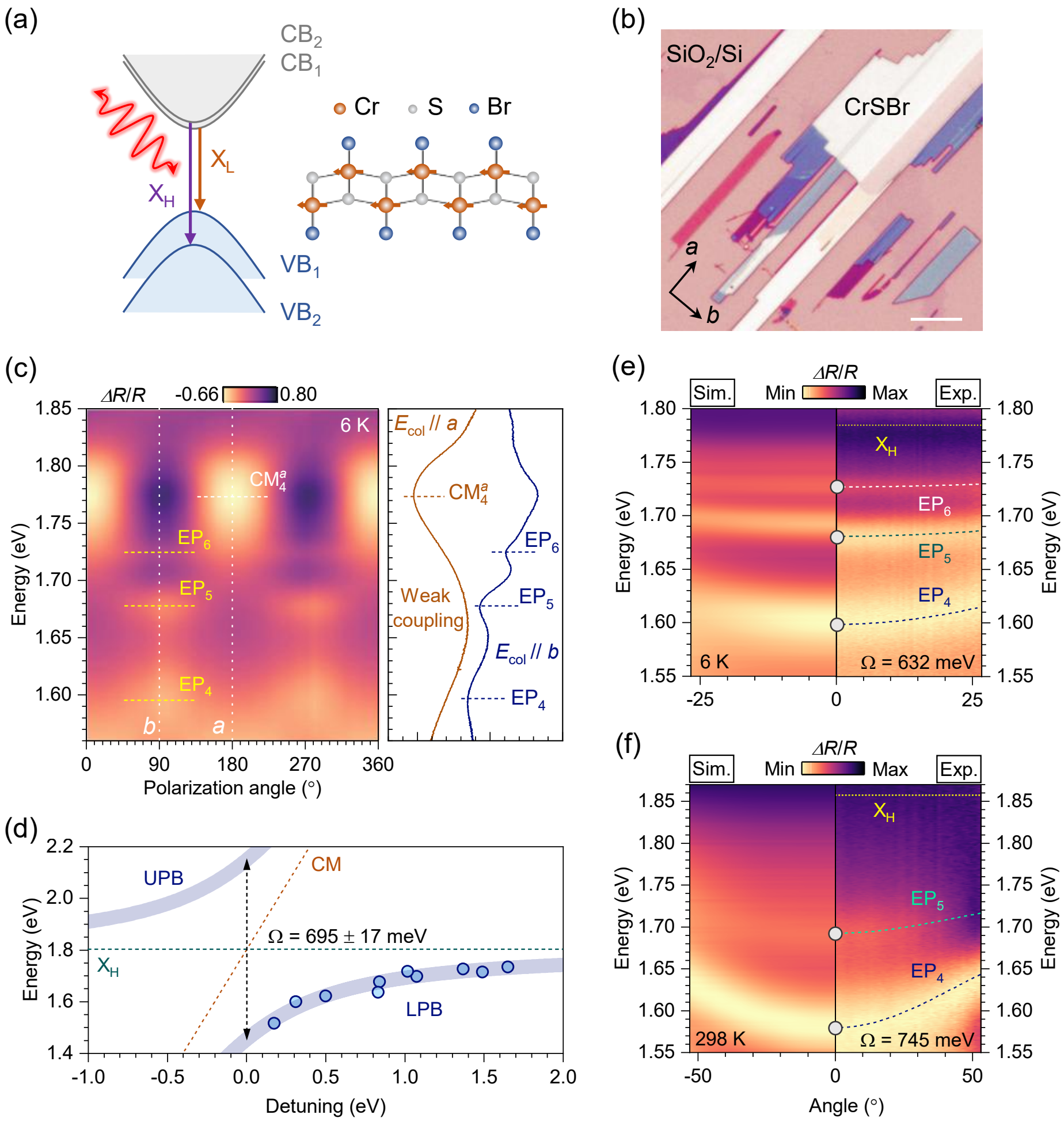

FIG. 1. (a) Schematic of $X_H$ exciton and $X_L$ exciton transition. CB and VB represent the conduction band and valence band, respectively. (b) Optical image of exfoliated CrSBr crystals on $SiO_2$/Si. Black arrows represent the *a*-axis and *b*-axis. Scale bar is 5 μm. (c) Left panel: differential reflectance spectra as a function of polarization of reflected light of the 354 nm-thick CrSBr crystal at 6 K. White dashed lines represent polarization along the *a*-axis and *b*-axis. Right panel: detailed reflectance spectra polarized along the *b*-axis (navy, strong coupling) and *a*-axis (brown, weak coupling). (d) Energy-detuning dispersion of CrSBr crystals with different thicknesses. The detuning energy is equal to the cavity mode CM energy minus $X_H$ exciton energy. Dots: experimental data at normal incidence. Navy curves: fitting results using the coupled harmonic oscillator model, giving an average Rabi splitting energy of 695 meV. UPB (LPB) stands for the upper (lower) polariton branch. (e, f) Simulated (left panel) and experimental (right panel) angle resolved reflectance

spectra of the 354 nm-thick CrSBr crystal at 6 K (e) and 298 K (f). Dotted curves and dashed curves denote the $X_H$ exciton and lower exciton-polariton branches. EP and CM represent the exciton-polariton and cavity mode, and subscripts indicate the mode number and polarization direction.

Figs. 1(d) and 1(e) presents angle-resolved reflectance spectra for a 354 nm-thick CrSBr crystal taken at 6 K and 298 K. The angle-resolved spectrum at 6 K exhibits three curved dispersions, well fitted using an exciton-polariton model with a Rabi splitting energy of 632 meV. The normalized coupling strength of 0.177 reveals the demonstration of ultrastrong exciton-photon coupling, dramatically exceeding that of the reported self-hybridized exciton-polariton systems. The larger dispersion curvature and increased linewidth in the lower branches indicate a reduction in effective mass and an increase in dissipation with the mixing of more photons. Due to phonon-assisted exciton linewidth broadening, only two exciton-polariton branches are observed at 298 K, with a fitted Rabi splitting energy of 745 meV. Further, the simulated angle-resolved reflectance spectra using finite-difference time-domain method are closely aligned with experimental observations.

The behaviors of $X_H$ excitons and corresponding exciton-polaritons are highly dependent on the magnetic order and can be effectively tailored by an applied external magnetic field. At 6 K, reflectance spectra at normal incidence as a function of the in-plane magnetic field ($B$, parallel to the easy $b$-axis) show that all exciton-polariton branches maintain the constant energies when $B$ is less than 0.35 T, followed by an abrupt redshift between 0.35-0.45 T, after which the energies stabilize again as $B$ exceeds 0.45 T [Figs. 2(a) and S7 [42]]. This sudden redshift arises from changes in interlayer electronic hybridization due to spin flipping, accomplished by the transition from AFM across IM to FM order [31]. Interestingly, the Rabi splitting energy also sharply decreases from 632 meV to 535 meV across the spin flipping transition. In contrast, the Rabi splitting energy for the coupling of $X_L$ excitons and photons is almost independent of the magnetic order (see Fig. S8 [42]). Therefore, the resulting modulation of light velocity and effective mass of exciton-polaritons, 100 meV below the $X_H$ exciton energy, respectively reach 10.3% and 26.1%, 1-2 orders of magnitude higher than those of $X_L$ exciton (see Fig. S9 and Note S2 [42]). The giant magnetic modulation of the Rabi splitting energy is due to the fact that the wavefunction of $X_H$ excitons exhibits a larger inter-site character, involving interactions between multiple Cr atomic sites in space, where electrons and holes may be located on different Cr atoms [39,43]. Therefore, $X_H$ excitons are more sensitive to changes in interlayer interactions between Cr atoms during the spin flipping, leading to the exciton wavefunction delocalization and the reduction of the Rabi splitting energy. In contrast, the contribution of local Cr atomic sites plays a crucial role in the wavefunction of $X_L$ excitons, resulting in negligible wavefunction delocalization and thus Rabi splitting energy tuning by magnetic fields. Further, the fitted energy redshift of $X_H$ excitons is ~78 meV (see Fig. S7 [42]), nearly five times that of $X_L$ excitons (~16 meV, see Fig. S8 [42]), reflecting dramatic field-induced band structure change associated with the $X_H$ exciton transition. The global vacuum energy (the sum of vacuum photon energy and exciton energy [44]) in the ultrastrong coupling system undergoes a variation (depending on coupling strength) compared with the uncoupled system, so that the directly measured energy shift of $X_H$ excitons in a few-layer sample without exciton-polaritons, ~88 meV (see Fig. S10 [42]), exhibits a slight difference from the fitted value (~78 meV) using the strong coupling model. When the temperature increases to 80 K, similar switching behaviors in Rabi splitting energy, $X_H$ exciton-polariton energy, and $X_H$ exciton energy are observed [Figs. 2(b) and S7, [42]]. The critical magnetic field for the spin flipping transition is lower (~0.15-0.20 T), above which a slight energy redshift still occurs. These results are due to thermal fluctuations weakening the pristine AFM coupling and field-induced FM coupling [37]. At 150 K (above the Néel temperature of 132 K for bulk CrSBr), CrSBr transforms into a paramagnetic state, and the magnetic response of both $X_H$ excitons and corresponding exciton-polaritons vanishes completely [Figs. 2(c) and S7, [42]].

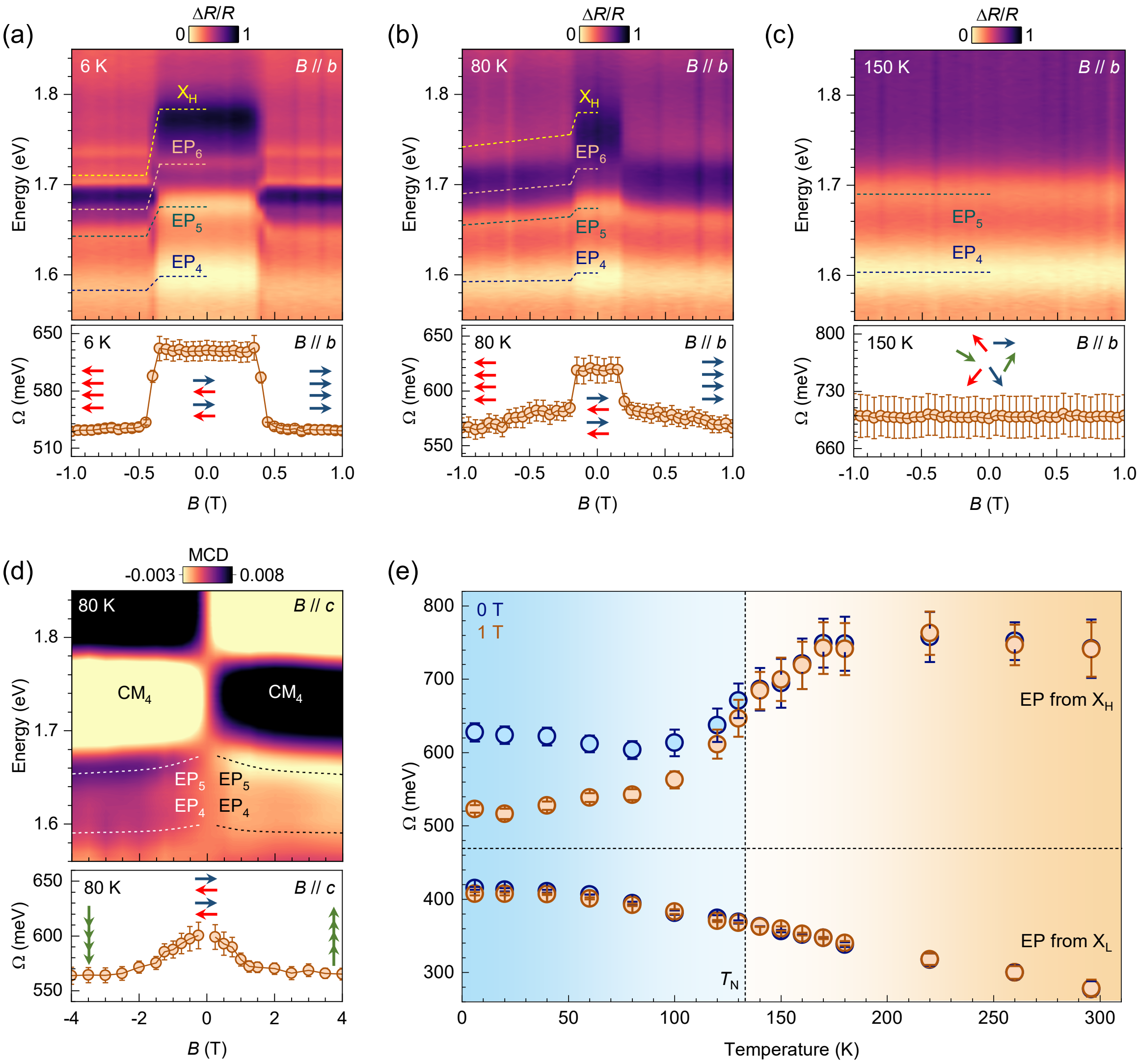

FIG. 2. (a-d) Upper panel: magnetic field dependence of differential reflectance spectra at 6 K (a), 80 K (d), 150 K (g) and MCD spectra at 80 K (j) of the 354 nm-thick CrSBr crystal. Lower panel: Magnetic field dependence of fitted Rabi splitting energies. (e) Temperature dependence of the Rabi splitting energy under $B$ = 0 T (navy) and $B$ = 1 T (brown). The data above (below) the horizontal dashed lines correspond to exciton-polaritons formed by $X_H$ excitons ($X_L$ exciton). The vertical dashed line represents the Néel temperature.

MCD spectroscopy was employed to further verify the magnetic tuning of ultrastrong coupling. Fig. 2(d) presents the out-of-plane magnetic field (parallel to the hard *c*-axis) dependence of MCD spectra at 80 K. The exciton-polariton branches exhibit a continuous redshift with increasing the magnetic field, consistent with the spin canting process. The redshift rate slows down significantly at $B > 1$ T, accompanied by saturation of the MCD intensity growth (see Fig. S11 [42]), highlighting the critical magnetic field strength for the AFM-FM phase transition.

Similarly, fitted Rabi splitting energy and $X_H$ exciton energy also continuously undergo a reduction and gradually stabilize beyond $B$ of 1 T [Figs. 2(d) and S7 [42]]. The consistency between the reflectance and MCD spectra at the same temperature of 80 K, across both AFM and FM phases, verifies the reliability of the data.

The temperature dependence of reflectance spectra was collected to unlock the evolution of exciton-polaritons influenced by excitons interacting with incoherent magnons and phonons (see Fig. S12 [42]). As $B = 0$ T, the Rabi splitting energy monotonically decreases below 100 K, and transform to increase between 100 and 200 K. As $B = 1$ T (above the critical field for the AFM-FM transition), an accelerated increase occurs below 200 K, resulting in a continuous reduction of field-induced tuning of Rabi splitting decreases from ~100 meV (~16%) at 6 K to around zero above 140 K [Fig. 2(e)]. Specifically, the temperature dependent behaviors of the Rabi splitting energy closely mirror the polariton energy and the fitted $X_H$ exciton energy, all of which can be well described by the exciton-magnon coupling and exciton-phonon coupling theory (see Fig. S13, Note S3 and Note S4 [42]), suggesting strong spin and phonon correlated exciton-photon coupling. The blueshift of the $X_H$ exciton energy below 200 K may be due to a positive temperature slope of the gap driven by the thermal expansion term [45]. There are still several meV magnetic tuning amounts above 140 K, possibly due to the existing short-range correlations above the Néel temperature [35]. Moreover, the ratio of the Rabi splitting energy to the $X_H$ exciton energy indicates the robust ultrastrong coupling up to room temperature (see Fig. S14 [42]). As a comparison, the Rabi splitting energy of exciton-polaritons formed by $X_L$ excitons can hardly be tuned by an external magnetic field.

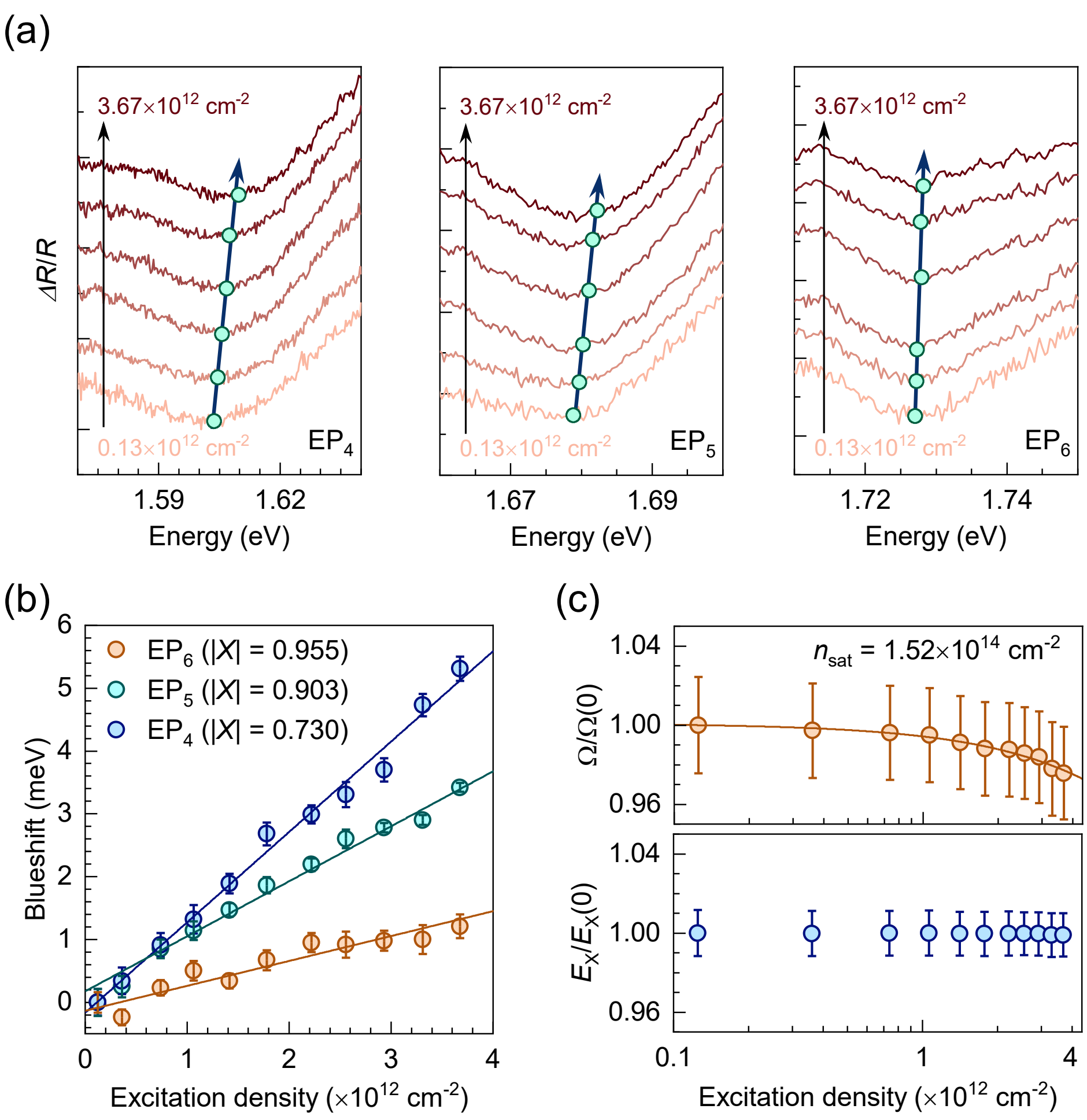

FIG. 3. (a, b) Excitation density dependence of reflectance spectra (a) and blueshift (b) for three exciton-polariton branches in the 354 nm-thick CrSBr crystal. The central energies of exciton-polaritons are marked with cyan circles in (a). (c) Fitted Rabi splitting energy (upper panel) and $X_H$ exciton energy (lower panel) relative to the Rabi splitting energy and $X_H$ exciton energy at the lowest excitation density as a function of excitation density according to (a).

Probing and manipulating polariton nonlinearity is crucial in accessing many-body quantum phenomena and quantum optical devices. We characterize the nonlinear response of exciton-polaritons by performing excitation density dependent reflectance measurements. Fig. 3(a) shows the representative reflectance spectra at normal incidence of three exciton-polariton branches in the 354 nm-thick crystal, and the polariton energy is obtained by fitting the reflectance dips using a Lorentzian function (see Fig. S15 [42]). As the excitation density increases, all branches exhibit a linear blueshift, and the magnitude of the blueshift gradually increases as the exciton component decreases [Fig. 3(b)]. Within the excitation density range of interest, the fitted $X_H$ exciton energy remains almost unchanged, reflecting the suppressed long-range excitonic interactions [Fig. 3(c)]. Similar behavior has been reported in other zero-dimensional and one-dimensional exciton systems with highly confined wavefunctions [13,34]. Due to the Pauli exclusion principle (phase-space filling effect), the Rabi splitting energy saturates as the excitation density approaches the fitted saturation exciton density of $1.52 \times 10^{14}$ $cm^{-2}$ (see Note S5, [42]). Considering the anisotropic dielectric function and the elliptical distribution of the exciton wavefunction, the predicted $X_H$ exciton Bohr radius along the *b*-axis and *a*-axis is 0.90 nm and 0.25 nm, respectively. The consistency of these results indicates that the strongly localized nature of $X_H$ excitons leads to the extremely large exciton oscillator strength and thus the occurrence of the ultrastrong coupling.

The magnetic field (parallel to the easy *b*-axis) effect on polariton nonlinearity has been further studied. Figs. 4(b) and 4(e) shows the excitation density-dependent reflectance spectra for two exciton-polariton branches at $B = 0$ T (AFM phase), 0.375 T (IM phase), and 1 T (FM phase). Compared to $B = 0$ T, both branches at $B = 1$ T exhibit a similar blueshift. However, at $B = 0.375$ T, the blueshift at excitation density of $5.1 \times 10^{12}$ $cm^{-2}$ of mode number $m = 4$ branch, increases to 10 meV, twice that at $B = 0$ T [Fig. 4(c)], and the $m = 5$ branch exhibits a 5 meV redshift, in contrast to the 3 meV blueshift at $B = 0$ T [Fig. 4(f)]. We attribute such results to additional long-range interaction mediated by magnons in the IM phase. The stronger exciton-magnon coupling enhances the attractive exciton-exciton interaction when the spin flipping begins [39], resulting in an energy redshift of $X_H$ excitons and $m = 5$ branch with a large exciton fraction. Meanwhile, the magnon-assisted long-range interactions lead to a decrease in exciton saturation density, which accelerates the Rabi splitting energy reduction and thereby the blueshift of $m = 4$ branch. Short-range exciton-exciton interactions and phase-space filling are primarily responsible for the polariton nonlinearity in the conventional systems, requiring exciton wavefunctions overlapping and filling phase-space at high excitation densities. Due to the strongly localized exciton wavefunctions in CrSBr, these two channels are suppressed, revealed by the unchanged exciton energy and weak Rabi splitting energy saturation. However, due to the strong coupling between excitons and magnons, as well as the spatial extension of spin waves, exciton-exciton interactions can occur over long ranges *via* magnons as mediators. In the AFM and FM phases, spin waves mainly retain their in-plane (2D-like) characteristics [Fig. 4(a)] due to weak interlayer exchange interactions in the vdW CrSBr [44]. In the IM phase, incomplete alignment of spins across different layers leads to non-zero interlayer coupling, which facilitates the formation of interlayer spin waves (3D-like) and thus both magnon-mediated long-range attractive interactions and the reduction of the coupling strength and consequently strongly affects the polariton nonlinearity.

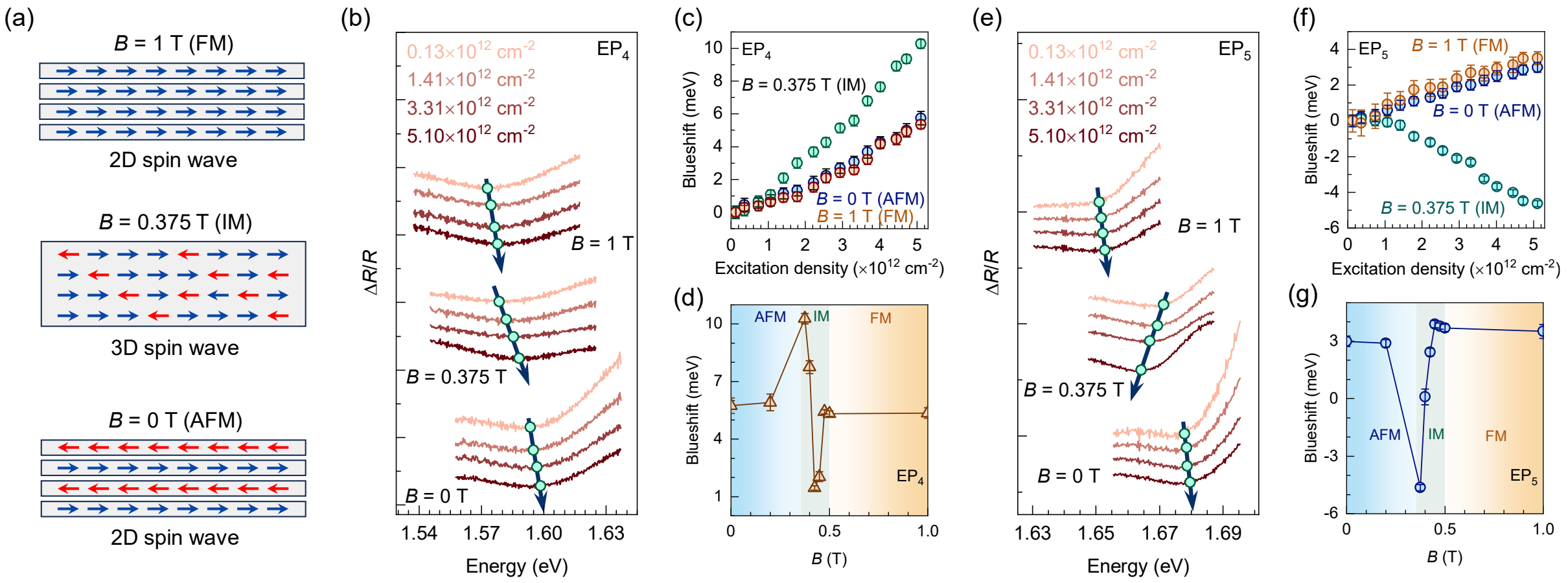

FIG. 4. (a) Schematic of spin ordering disruption resulting in the transformation of spin wave from 2D to 3D. (b, c, e, f) Excitation density dependence of reflectance spectra (b, e) and blueshift (c, f) for $m = 4$ (b, c) and $m = 5$ polariton branches (e, f) in the 354 nm-thick CrSBr crystal under $B = 0$ T (FM phase), 0.375 T (IM phase), and 1 T (FM phase). Colored arrows show the possible orientation of magnetizations. The central energies of exciton-polaritons are marked with cyan circles in (b) and (e). (d, g) Magnetic field dependence of blueshift at $5.1 \times 10^{12}$ cm$^{-2}$ for $m = 4$ (d) and $m = 5$ branch (g). Negative blueshift means redshift.

We have added data at other magnetic fields and confirmed that the energy shift of the exciton-polaritons by varying excitation density can be continuously controlled in the IM phase [Figs. 4(d) and 4(g)]. The shift of $m = 4$ ($m = 5$) branch can be tuned from a redshift of 5 meV to a blueshift of 4 meV (from a blueshift of 1 meV to 10 meV). Since spin flipping occurs rapidly within a small magnetic field range in our configurations, the polariton nonlinearity dramatically increases and then quickly returns, different from the gradually evolving exciton-exciton interaction observed during spin canting in a recent work [39]. The demonstrated sensitive and customizable polariton nonlinearity provides a promising approach for developing flexible nonlinear optical switches and quantum simulators.

*Conclusions*—In this study, we demonstrate strong coupling between magnetically dressed $X_H$ excitons and photons in the vdW CrSBr crystals with a Rabi splitting energy up to 745 meV and operation temperature from 6 to 298 K. We find that the Rabi splitting energy and nonlinearity of exciton-polaritons formed by $X_H$ excitons are extensively tuned by an applied in-plane magnetic field within 1 T, and the magnetic tunability is 1-2 orders of magnitude larger than that of CrSBr $X_L$ excitons and/or other vdW semiconductor excitons, which is due to significant exciton wavefunction delocalization and magnon-assisted attractive polariton interactions and phase space filling. These results provide valuable insights into the underlying physics and practical applications involving magnetic exciton-polaritons.

**ACKNOWLEDGMENTS**

Q.Z. acknowledges funding support from National Natural Science Foundation of China (U23A2076, 51991340, 52072006) and Beijing National Natural Science Foundation (JQ21004). C.S. acknowledges funding support from National Key Research and Development Program of China (2022YFF1201900).